\documentclass[aps,twocolumn,showpacs,showkeys,nofootinbib]{revtex4}
\usepackage{epsfig}
\usepackage{amsmath}
\usepackage{amsfonts}
\usepackage{amssymb}
\usepackage{graphicx}
\usepackage{colordvi}
\begin{document}

\title{A physical interpretation of the cubic map coefficients describing the electron cloud evolution}

\author{T. Demma$^1$\footnote{e-mail address: theo.demma@infn.lnf.it}, S. Petracca$^2$ $^3$\footnote{e-mail address: petracca@sa.infn.it}, A. Stabile$^2$\footnote{e-mail address: arturo.stabile@gmail.com}}

\affiliation{$^1$INFN-LNF, Frascati, Italy\\$^2$Dipartimento di Ingegneria, Universita'
del Sannio, Corso Garibaldi, I - 80125 Benevento, Italy\\$^3$INFN Salerno, Italy}

\begin{abstract}
The \emph{Electron Cloud} (ecloud), an undesirable physical
phenomena in the accelerators, develops quickly as photons
striking the vacuum chamberwall knock out electrons that are then
accelerated by the beam, gain energy, and strike the chamber
again, producing more electrons. The interaction between the
electron cloud and a beam leads to the electron cloud effects such
as single- and multi-bunch instability, tune shift, increase of
pressure and particularly can limit the ability of recently build
or planned accelerators to reach their design parameters. We
report a principal results about the analytical study to
understanding a such dynamics of electrons.
\end{abstract}

\pacs{XX; XX; XX}
\keywords{XXXXX}
\maketitle

\section{Introduction}\label{sec:intro}

The generation of a quasi-stationary electron cloud inside the
beam pipe through beam-induced multipacting has become an area of
intensive study. The analysis performed so far was based on very
heavy computer simulations (ECLOUD \cite{zim}) taking into account
photoelectron production, secondary electron emission, electron
dynamics, and space charge effects providing a very detailed
description of the electron cloud evolution.

In \cite{iriso1} has been shown that, for the typical parameters
of \emph{Relativistic Heavy Ion Collider} (RHIC), the evolution of
the electron cloud density can be followed from bunch to bunch
introducing a "cubic map" of the form:

\begin{eqnarray}\label{eq:cubic_map}
n_{m+1}\,=\,\alpha\,n_m+\beta\,{n_m}^2+\gamma\,{n_m}^3
\end{eqnarray}

where $n_m$ is the average electron cloud density after $m$-th
passage of bunch, $n_{m+1}$ is the one after $(m+1)$-th passage.
The coefficients $\alpha$, $\beta$, $\gamma$ are the parameters
extrapolated from simulations, and are functions of the beam
parameters and of the beam pipe features. The average longitudinal
electron density as function of time grows exponentially until the
space charge due to the electrons themselves produces a saturation
level. Once the saturation level is reached the average electron
density does not change significantly. The final decay corresponds
to the succession of the empty bunches.

\begin{figure}[htbp]
\includegraphics[scale=.5]{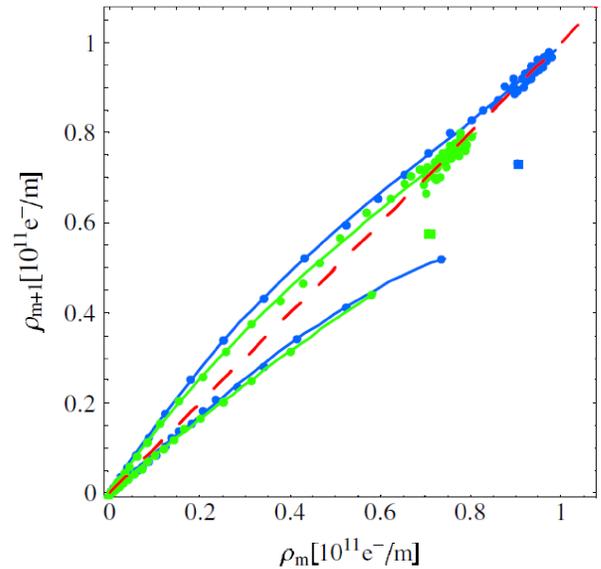}\\
\caption{Time evolution of the electron density (green line)
computed with ECLOUD. The black dots mark the average electron
density between two consecutive bunches.} \label{fig_demma_1}
\end{figure}

A such map approach has been proved, by numerical simulations,
reliable also for \emph{Large Hadron Collider} (LHC) \cite{demma}.
The most important outcomes of map formalism for LHC (and
generally for any accelerator) are summarizable as follows. In
Fig. \ref{fig_demma_1} one can see that the bunch-to-bunch
evolution contains enough information about the build-up or the
decay time, although the details of the line electron density
oscillation between two bunches are lost.

\begin{figure}[htbp]
\includegraphics[scale=.5]{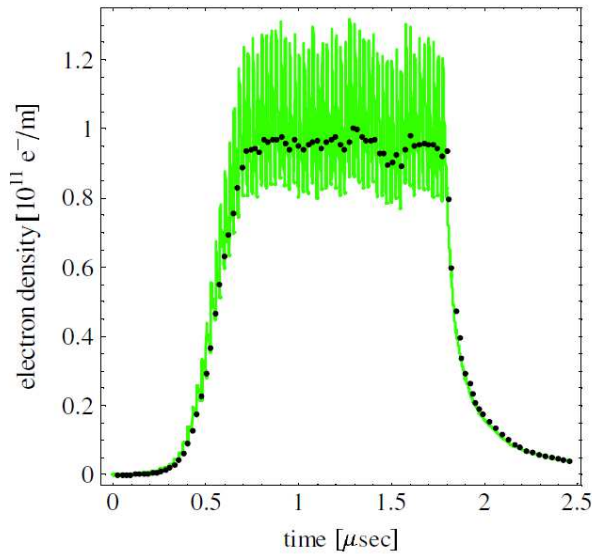}\\
\caption{Average longitudinal electron density after the passage
of bunch as a function of longitudinal electron density $n_{m+1}$
before the passage of $m$-th bunch, for different bunch
intensities (green $N_b\,=\,8*10^{10}$, blue
$N_b\,=\,16*10^{10}$). The lines correspond to cubic fits applied
to the average bunch to bunch points. The red line corresponds to
the identity map $n_{m+1}\,=\,n_m$. Points above this line
describe the initial growth and saturation of the bunch-to-bunch
evolution of the line electron density, those below describe the
decay. The black line represents the cubic fit to the points
corresponding to the first empty bunches.} \label{fig_demma_2}
\end{figure}

Fig. \ref{fig_demma_2} shows the behavior of the average electron
density $n_{m+1}$, after the passage of $(m+1)$-th bunch, as
function of the average electron density as $n_{m}$ for different
bunch intensities.

The points in Fig. \ref{fig_demma_2} show the average electron
cloud density between two bunches using results from ECLOUD, the
lines are cubic fits to these points. The physical dynamics of
electron cloud is explained as follows: starting with a small
initial linear electron density, after some bunches the density
takes off and reaches the corresponding saturation line
($n_{m+1}\,=\,n_m$, red line) where the space charge effects due
to the electrons in the cloud itself take place. In this
situation, all the points (corresponding to the passage of full
bunches) are in the same spot. The justification of the three
terms in eq. (\ref{eq:cubic_map}) is explained as a consequence of
the linear growth (this term has to be larger than unity in case
of electron cloud formation), a parabolic decay due to space
charge effects (this term has to be negative to give concavity to
the curve $n_{m+1}$ vs $n_m$), and a cubic term corresponding to
perturbations. Neglecting the point corresponding to the electron
cloud density after the first empty bunch, the longitudinal
electron density follows a similar decay independently of the
initial value of the saturated line electron density. The points
corresponding to the first empty bunches coming from different
saturation values lie on a general curve (black curve in Fig.
\ref{fig_demma_2}). Thus the electron density build up for a given
bunch intensity is determined by a cubic form, while decay is
described by two different cubic forms, one corresponding to the
first empty bunch, and a second to the rest.

Even though the behavior of the map coefficients from simulations
has been compared with the experiments always this is not well
understood and the determination of their values is purely
empirical. An analytical expression for coefficient $\alpha$
(linear coefficient) in the case of drift space has been found by
understanding the dynamics that governs weak cloud behavior
\cite{iriso1}.

Let us consider $N_{el,m}$ quasi-stationary electrons
gaussian-like distributed in the transverse cross-section of the
beam pipe. The $(m+1)$-th bunch accelerates the $N_{el,m}$
electrons initially at rest to an energy $\mathcal{E}_g$. After
the first wall collision two new jets are created: the
backscattered one with energy $\mathcal{E}_g$ and proportional to
$\delta_r(\mathcal{E}_g)$, and the "true secondaries" (with energy
$\mathcal{E}_0\,\sim\,5\,eV$) proportional to
$\delta_t(\mathcal{E}_0)$. The functions $\delta$'s (or $SEY$:
\emph{Secondary emission yield}) give the ratio of emitted
secondary electrons per incident electron. The sum of all these
jets becomes the number of surviving electrons $N_{el,m+1}$ (see
section \ref{sec:coefficient}), and we have

\begin{eqnarray}\label{linear_coeff}
\alpha\,=\,\frac{N_{el,m+1}}{N_{el,m}}
\end{eqnarray}

In this paper we derive, by assuming the previous theoretical
expression for linear coefficient ($\alpha$) \cite{iriso1}, a
simple approximate formula for the quadratic coefficient
($\beta$), which determines the saturation of the cloud due to
space charge, in the electron cloud density map, under the
assumptions of round chambers and free-field motion of the
electrons in the cloud. The coefficient depends on the bunch
parameters, and can be simply deduced from ecloud simulation codes
modelling the involved physics in full detail. Results are
compared with simulations for a wide range of parameters governing
the evolution of the electron cloud.

In the section \ref{density} we calculate the electronic density
of saturation by imposing a gaussian-like distribution for the
space charge and requiring the presence of energy barrier near to
wall of chamber. In the section \ref{sec:coefficient} we report
the calculus of linear coefficient and we compute the calculus of
quadratic term. Once calculated saturation we pass to estimate
theoretically the coefficient $\beta$. We conclude (section
\ref{sec:conclu}) with the comparison with respect to outcomes of
numerical simulations obtained using ECLOUD \cite{zim}. In the
table \ref{tab:tab1} we report all parameters used for the the
calculations.

\begin{table}[!h]
\centerline{
\begin{tabular}{||l|c|c|r||}
\hline\hline
Parameter& Quantity & Unit & Value \\
        \hline
        Beam pipe radius & $b$ & m & $.045$ \\
        Beam size & $a$ & m & $.002$ \\
        Bunch spacing & $s_b$ & m & $1.2$\\
        Bunch length & $h$ & m & $.013$\\
        Energy for $\delta_{max}$ & $\mathcal{E}_{0,max}$ & eV & $300$\\
        - & $\mathcal{E}_r$ & eV & $60$\\
        Particles per bunch & $N_b$ & $10^{10}$& $4\,\div\,9$\\
        SEY (max) & $\delta_{max}$ & - & $1.7$\\
        SEY ($\mathcal{E}\rightarrow0$) & $\delta_0$ & - & $.7$\\
        SEY ($\mathcal{E}\rightarrow\infty$) & $\delta_{\infty}$ & - & $.15$\\
        - & $\zeta$ & - & 1.83\\
\hline\hline
\end{tabular}}
\caption{Input parameters for analytical estimate and ECLOUD
simulations.} \label{tab:tab1}
\end{table}

\section{Steady-state: Electronic density of Saturation}\label{density}

In the chamber we have two groups of electrons belonging
to cloud: primary photo-electrons generated by the synchrotron
radiation photons and secondary electrons generated by the beam
induced multi-pactoring. Electrons in the first group generated at
the beam pipe all with the radius $b$ interact with the parent
bunch and accelerated (by a short bunch) to the velocity
$v/c\,=\,2\bar{N}_br_e/b$, where $r_e$ is the classical electron
radius and $\bar{N}_b$ is the efficacy value of bunch population.
Since we are in presence of trains of finite length beams to consider
a similar analysis to costing hypothesis it should replace the value of bunch population
with its spatial or temporal average. In fact we have

\begin{eqnarray}
\bar{N}_b\,=\,\frac{h}{h+s_b}N_b
\end{eqnarray}
where $s_b$ is the bunch spacing and $h$ is the length of bunch.
Electrons in the second group, generally, miss the parent bunch and move from the
beam pipe wall with the velocity $v/c\,=\,\sqrt{2\mathcal{E}_0/mc^2}$ until
the next bunch arrives. The velocity is defined by the average energy $\mathcal{E}_0$
of the secondary electrons and, at high $\bar{N}_b$,
is smaller than velocity of the first group. The process of the
cloud formation depends, respectively, on two parameters:

\begin{eqnarray}
k\,=\,\frac{2\bar{N}_br_eh}{b^2}
\end{eqnarray}
\begin{eqnarray}
\xi\,=\,\frac{h}{b}\sqrt{\frac{2\mathcal{E}_0}{mc^2}}
\end{eqnarray}
These parameters are the distance (in units of $b$) passed by
electrons of each group before the next bunch arrives. At low
currents, $k\,<<\,1$, electron interact with many bunches before
it reaches the opposite wall. In the opposite extreme case,
$k\,>\,2$, all electrons go wall to wall in one bunch spacing. The
transition to the second regime can be expected , therefore,
$k\,\sim\,1$ where the cloud is quite different than it is at low
currents. For $k\,>\,1$, secondary electrons are confined within
the layer $\xi\,<\,r/b\,<\,1$ at he wall and are wiped out of the
region $0\,<\,r/b\,<\,\xi$ close to the beam by each passing
bunch. This makes the range of parameters ($k\,>\,1$ and
$2-k\,<\,\xi\,<1$) quite desirable to suppress the adverse effects
of the e-cloud on the beam dynamics \cite{eifets}.

The condition of neutrality implies that secondary electrons
remain in the cloud for a time long enough to affect the secondary
electrons generated by the following bunches. In other words, the
condition of neutrality and the quasi-steady equilibrium
distribution of the electron cloud are justified only for small
$k$. It is not the case at the high currents. In this case, all
primary photo-electrons disappear just in one pass. The secondary
electrons are produced with low energy $\mathcal{E}_0$ and are
locked up at the wall. The density of the secondary electrons
grows until the space-charge potential of the secondary electrons
is lower than $\mathcal{E}_0$. The saturation condition can be
obtained by requiring that the potential barrier is greater than
electron energy in the point $r/b\,=\,1-\xi$

\begin{eqnarray}\label{condition}
-e\,V(1-\xi)\,\sim\,\mathcal{E}_0
\end{eqnarray}
where $V$ is the electric potential generated by the bunch and
electron cloud and $-e$ is the electron charge. Here it needs to
calculate the electric potential by assuming some model for the
electronic density.

\begin{figure}[htbp]
\includegraphics[scale=1]{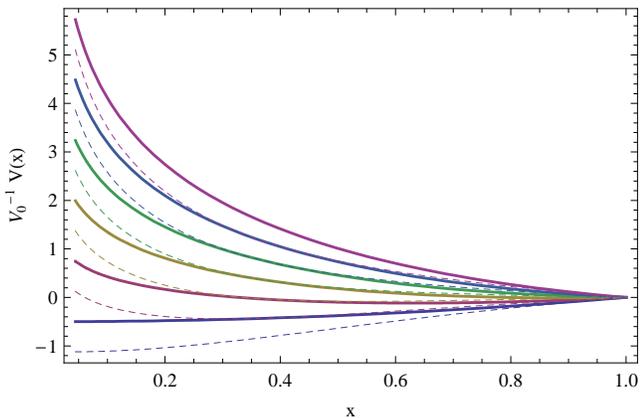}\\
\caption{Plot of $V_0^{-1}V(x)$, (\ref{radialpotential}) and
(\ref{potentialuniform})), in the case of uniform (solid lines)
and gaussian (dashed lines) electronic distribution for
$g\,=\,0\,\div\,2$, $\tilde{a}\,=\,.04$, $\tilde{r}_0\,=\,0$,
$\tilde{\sigma}\,=\,.3$.} \label{plotpotential}
\end{figure}

Our system is composed by a chamber with radius $b$, a bunch with
radius $a$ and length $h$, an electron cloud with density $\rho$.
Let us consider a electron cylindric distribution with a radial
gaussian density centered in $r_0$ as follows

\begin{eqnarray}\label{radialdensity_1}
\rho(r)\,=\,\rho_0e^{-\frac{(r-r_0)^2}{2\sigma^2}}
\end{eqnarray}
where $\rho_0$ is fixed by the condition

\begin{eqnarray}
2\pi h\int_a^b\rho(r)rdr\,=\,-N_{el}\,e
\end{eqnarray}
where $N_{el}$ is the total number of electrons in the volume $\pi h(b^2-a^2)$. The electric field in the chamber is

\begin{eqnarray}\label{electricfield}
\vec{E}\,=\,&&\biggl\{\frac{\bar{N}_b\,e}{2\pi\epsilon_0h}\frac{1}{r}-\frac{N_{el}\,e}{2\pi\epsilon_0h}
\frac{1}{\int_a^b
e^{-\frac{(y-r_0)^2}{2\sigma^2}}y
dy}\times\nonumber\\\nonumber\\
&&\times\frac{\int_a^re^{-\frac{(y-r_0)^2}{2\sigma^2}}y
dy}{r}\biggr\}\hat{r}
\end{eqnarray}
The electric potential $V(r)$ defined by the condition $V(b)\,=\,0$ is


\begin{eqnarray}\label{radialpotential}
V(r)\,=&&\int_r^b\vec{E}\cdot
d\vec{l}\,=\,\nonumber\\&&-\frac{\bar{N}_b\,e}{2\pi\epsilon_0h}\ln\frac{r}{b}-\frac{N_{el}\,e}{2\pi\epsilon_0h}
\frac{\int_r^b\,\frac{\int_a^y\,e^{-\frac
{(z-r_0)^2}{2\sigma^2}}z\,dz}{y}dy}{\int_a^b\,e^{-\frac{(y-r_0)^2}{2\sigma^2}}y\,dy}\,=\nonumber\\\nonumber\\&&-\frac{\bar{N}_b\,
e}{2\pi\epsilon_0h}\ln x-\frac{N_{el}\,
e}{2\pi\epsilon_0h}\frac{G(x)}{F(1)}\, =\nonumber\\\nonumber\\&&-V_0\biggl[g\ln
x+\frac{G(x)}{F(1)}\biggr]
\end{eqnarray}

where $F(x)\,=\,\int_{\tilde{a}}^x\,e^{-\frac
{(\tilde{y}-\tilde{r}_0)^2}{2\tilde{\sigma}^2}}y\,dy$,
$G(x)\,=\,\int_x^1\,\frac{F(y)}{y}dy$, $g\,=\,\bar{N}_b/N_{el}$,
$V_0\,=\,\frac{N_{el}\,e}{2\pi\epsilon_0h}$  and $x\,=\,r/b$,
$\tilde{a}\,=\,a/b$, $\tilde{r}_0\,=\,r_0/b$,
$\tilde{\sigma}\,=\,\sigma/b$. We note that in the case
$\sigma\,>>\,b$ (or $\tilde{\sigma}\,>>\,1$) and $r_0\,=\,0$ we
reobtain the uniform electron cloud and with $a\,\rightarrow\,0$
we must neglect the radial dimension of bunch with respect to one
of electron cloud. In fact in this case for eq.
(\ref{radialpotential}) we would have

\begin{eqnarray}\label{potentialuniform}
V_{ud}(r)\,=\,-V_0\biggl[g\ln x+\frac{1-x^2}{2}\biggr]
\end{eqnarray}
Obviously the potentials depend on $g$, the ratio of the densities
of the beam and of the cloud averaged over the beam pipe
cross-section. In Fig. \ref{plotpotential} we report the spatial
behavior of two potentials. The potential (\ref{potentialuniform})
has minimum at $r\,=\,r_m\,=\,b\sqrt{g}$ and is monotonic for
$g\,>\,1$ within the beam pipe. For $g\,<\,1$ it has minimum at
the distance $r_m\,<\,b$, and the condition $g\,=\,1$ defines the
maximum density. this is the well known condition of the
neutrality. The condition formulated in this form is, actually,
independent of the form of distribution. Similar behavior is found
also for the gaussian distribution density and is compared with
respect to previous one (Fig. \ref{plotpotential}).

\begin{figure}[htbp]
\includegraphics[scale=1]{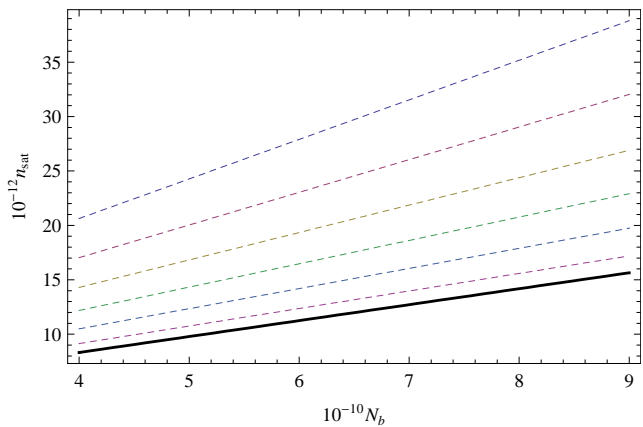}\\
\caption{Plot of electronic densities of saturation $n_{sat}$
vs $N_b$, (\ref{nsatgauss}) and (\ref{nsatunif})), in the case of
uniform (solid line) and gaussian (dashed lines) electronic
distribution for $\tilde{a}\,=\,.04$, $\tilde{r}_0\,=\,0$,
$\tilde{\sigma}\,=\,.3$ and $p\,=\,2\,\div\,3$}
\label{plotdensitysaturation}
\end{figure}

By imposing the condition eq. (\ref{condition}) we find the
critical number (saturation condition) of electrons in the chamber

\begin{eqnarray}
N_{el,sat}\,=\,\frac{2\pi\epsilon_0hF(1)\mathcal{E}_0}{e^2G(1-\xi)}-\frac{
F(1)\ln(1-\xi)}{G(1-\xi)}\bar{N}_b
\end{eqnarray}
while the average density of saturation is found by assuming that
electrons are confined in a cylindrical shell with inner radius
equal to $a$ and external radius to $r_0+p\,\sigma$ where $p$ is a
free parameter. So

\begin{eqnarray}\label{nsatgauss}
n_{sat}\,=\,\frac{N_{el,sat}}{\pi h b^2 [(\tilde{r}_0+p\,\tilde{\sigma})^2-\tilde{a}^2]}
\end{eqnarray}
where $p$ is a free parameter. In the case of uniform distribution
of electron cloud with a similar mathematical passages we find the
density of saturation

\begin{eqnarray}\label{nsatunif}
\bar{n}_{sat}\,=\,\frac{\bar{N}_{el,sat}}{\pi h b^2 [1-\tilde{a}^2]}
\end{eqnarray}
In the Fig. \ref{plotdensitysaturation} we show the behavior of
density of saturation (\ref{nsatgauss}) and (\ref{nsatunif}). It
is obvious in the case of a gaussian distribution of cloud we get
a estimate of density saturation greater than that of uniform
distribution. In fact, the same number of electrons occupies a
smaller volume (due to the Gaussian distribution).

\section{Analytical Determination of Coefficients}\label{sec:coefficient}

To compute the linear term we can assume that all the $N_{el,m}$
electrons, distributed in the transverse cross section of the beam
pipe, gain an energy $\mathcal{E}_g$ during the passage of the
bunch $m$. After the bunch passage, electrons are accelerated towards
the chamber wall
and have their first wall collision when two new jets are created: one with energy
$\mathcal{E}_g$ and $N_m\delta_r$ electrons, corresponding to
backscattered electrons; the second with low energy
and $N_m\delta_t$ electrons, corresponding to the true secondaries.
Before $(m+1)$-th bunch arrives, these two jets
perform several wall collisions, which in turn create more jets.
The contribution of all these jets becomes the number of surviving
electrons, $N_{m+1}$.
The reflected electrons travel across the beam pipe with energy
$\mathcal{E}_g$ and perform a number of collisions with the
chamber wall, between two consecutive bunches, that is:

\begin{eqnarray}
s\,=\,\frac{t_{sb}-t_f(\mathcal{E}_g)}{t_f(\mathcal{E}_g)},
\end{eqnarray}
where $t_{sb}$ is the time of bunch spacing,

\begin{eqnarray}
t_f(\mathcal{E}_g)\,=\,\frac{4b}{\pi \sqrt{2\mathcal{E}_g/m_e}}
\end{eqnarray}
is the average flight time and $\mathcal{E}_g$ is the energy of
electrons accelerated by bunch

\begin{eqnarray}
\mathcal{E}_g\,=\,m_ec^2\frac{r_eN_b}{\sqrt{2\pi}h}\biggl[\log\biggl(\frac{b}{1.05a}\biggr)-\frac{1}{2}\biggr]
\end{eqnarray}
Hence, the total number of reflected, high energy electrons at the
passage of $m+1$-th bunch is:

\begin{eqnarray}
N_{el|ref}\,=\,N_{el,m}{\delta_r}^s
\end{eqnarray}
The true secondaries electrons produced after the first wall
collision gives rise to a low energy jet ($\mathcal{E}_0$). For
this jet there is no distinction for the true secondaries and
reflected, since all are produced with the same energy. After the
$i-$th wall collision the number of surviving electrons is:

\begin{eqnarray}
N_{el|tot}\,=\,N_{el,m}\,\delta_t\,{\delta_r}^{i-1}\,{\delta_{tot}}^{k_i}
\end{eqnarray}
where
$\delta_{tot}\,=\,\delta_r(\mathcal{E}_0)+\delta_t(\mathcal{E}_0)$
and

\begin{eqnarray}
k_i\,=\,\frac{t_{sb}-it_f(\mathcal{E}_0)}{t_f(\mathcal{E}_0)},
\end{eqnarray}
is the number of collisions after the $i-$th collision. The low
energy electrons at the passage of $m+1$-th bunch is :

\begin{eqnarray}
N_{el|s}\,=\,N_{el,m}\delta_t\,\sum_{i=1}^s{\delta_r}^{i+1}{\delta_{tot}}^{k_i}
\end{eqnarray}
Finally the total number of survival electrons at $m+1$-th bunch
passage is obtained taking into account both the high and low
energy contributions:

\begin{eqnarray}
N_{el,m+1}\,=\,N_{el,m}\biggl[{\delta_r}^s+\delta_t\sum_{i=1}^s{\delta_r}^{i+1}{\delta_{tot}}^{k_i}\biggr]
\end{eqnarray}
and the linear term (\ref{linear_coeff}) can be written in the
form:

\begin{eqnarray}\label{linear_coeff_2}
\alpha\,=\,\frac{N_{el,m+1}}{N_{el,m}}\,=\,{\delta_r}^s+\delta_t{\delta_{tot}}^\eta\,\frac{{\delta_{tot}}^{s\eta}-
{\delta_r}^s}{{\delta_{tot}}^{\eta}-{\delta_r}^s}
\end{eqnarray}
where $\eta\,=\,\sqrt{\mathcal{E}_0/\mathcal{E}_g}$. The
expressions of $\delta$'s used are

\begin{eqnarray}
\left\{\begin{array}{ll}\delta_r(\mathcal{E})\,=\,\delta_\infty+(\delta_0-\delta_\infty)e^{-\frac{\mathcal{E}}{
\mathcal{E}_r}}
\\\\
\delta_t(\mathcal{E})\,=\,\zeta\,\delta_{max}\frac{\mathcal{E}}{\mathcal{E}_{0,max}}\frac{1}{\zeta-1+\frac{
\mathcal{E}^\zeta}{{\mathcal{E}_{0,max}}^\zeta}}
\end{array} \right.
\end{eqnarray}
where $\delta_\infty$, $\delta_0$, $\mathcal{E}_r$, $\zeta$,
$\delta_{max}$, $\mathcal{E}_{0,max}$ are the parameters of models
and their values are shown in Table \ref{tab:tab1}. The
coefficient $\beta$ can be found by imposing the saturation
condition of map (\ref{eq:cubic_map}):

\begin{eqnarray}\label{quadratic_coeff}
n_{sat}\,=\,\alpha\,n_{sat}+\beta\,{n_{sat}}^2\,\,\,\,\rightarrow\,\,\,\,\beta\,=\,\frac{1-\alpha}
{n_{sat}}
\end{eqnarray}
and the map (\ref{eq:cubic_map}) becomes

\begin{eqnarray}\label{map_quadratic}
n_{m+1}\,=\,\alpha\,n_m+\frac{1-\alpha}{n_{sat}}\,{n_m}^2
\end{eqnarray}

\begin{figure}[htbp]
\includegraphics[scale=1]{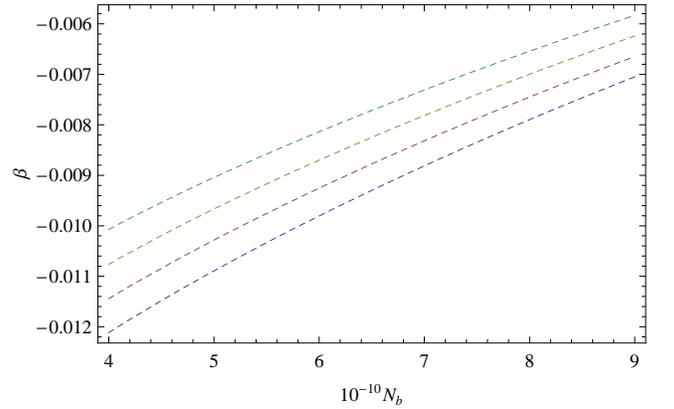}\\
\caption{Analytical prediction of coefficient $\beta$
(\ref{quadratic_coeff}) for values $\delta_{max}\,=\,1.4\div2$ and
$p\,=\,2$.} \label{plot_b_dmax}
\end{figure}

\begin{figure}[htbp]
\includegraphics[scale=1]{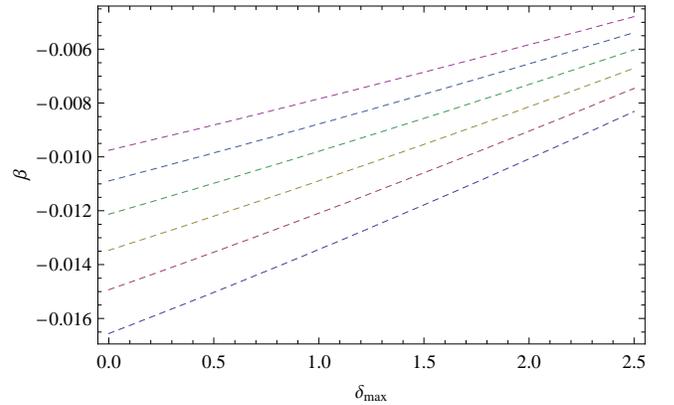}\\
\caption{Analytical prediction of coefficient $\beta$
(\ref{quadratic_coeff}) for values $N_b\,=\,(4\div9)*10^{10}$ and
$p\,=\,2$. } \label{plot_b_Nb}
\end{figure}

In Fig. (\ref{plot_b_dmax}), (\ref{plot_b_Nb}) we show the trends
of the coefficient (\ref{quadratic_coeff}) as a function of
$\delta_{max}$ for various values of bunch population and
viceversa.


\section{Conclusions}\label{sec:conclu}

\begin{figure}[htbp]
\includegraphics[scale=1]{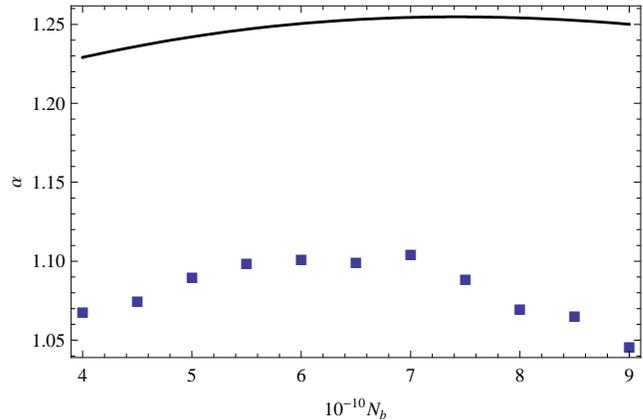}\\
\caption{Comparison between the analytical prediction (solid
line) of $\alpha$ (Eq. (\ref{linear_coeff_2})) and the simulation
(points) by ECLOUD code.} \label{plot_a}
\end{figure}

\begin{figure}[htbp]
\includegraphics[scale=1]{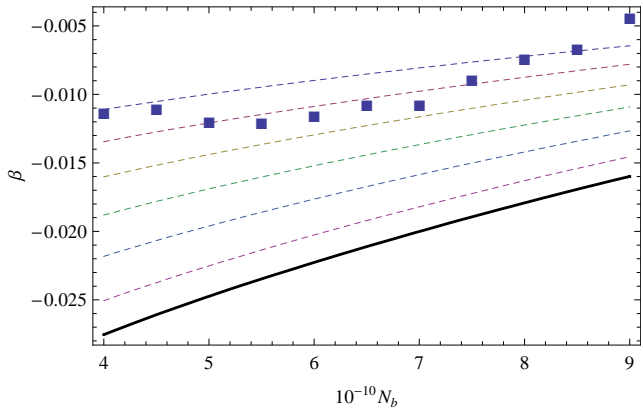}\\
\caption{Comparison of the quadratic coefficient $\beta$ (Eq.
(\ref{quadratic_coeff})) derived using ECLOUD simulations (points)
and using the analysis of previous sections (dashed lines) with
$p\,=\,2\,\div\,3$. The solid line is the result by assuming an
uniform density.} \label{plot_b}
\end{figure}

In Figs. (\ref{plot_a}) and (\ref{plot_b}) we report the
analytical behaviors and the outcomes of simulations (ECLOUD code)
of coefficients $\alpha$ and $\beta$ for values in table
\ref{tab:tab1}. It is important to note about the quadratic
coefficient we had a good agreement with simulations because in
our theoretical treatment we have a degree of freedom due to the
choice of charge distribution in the chamber. By requiring the
condition of saturation it needs to choose an average density that
can be as realistic as possible. A similar discourse was not
possible for the linear term because we started from the result of
Iriso \& Peggs (\cite{iriso2}) and then we built the working
hypothesis for the determination of $\beta$. This specification
allows us to evaluate the apparent discrepancy between the linear
and quadratic coefficient with respect to their results from
ECLOUD code.

\begin{figure}[htbp]
\includegraphics[scale=1]{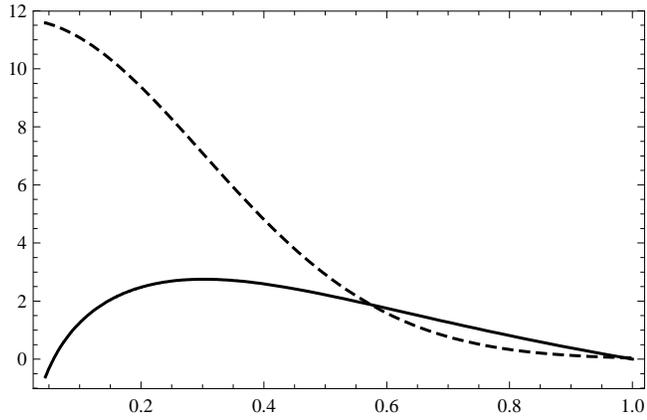}\\
\caption{Plot of electronic density,(\ref{radialdensity}),
$10^{-15}n(r)$ (dashed line) and of energy barrier
(\ref{radialenergy}), $10^{15}\mathcal{E}(r)$ (solid line) for
$\tilde{\sigma}\,=\,.3$,
$\tilde{r}_0\,=\,0$.}\label{plot_density_energy}
\end{figure}

The radial profile of electric density $n(r)$, from
(\ref{radialdensity_1}), is

\begin{eqnarray}\label{radialdensity}
n(r)\,=\,\frac{\rho(r)}{-e}\,=\,\frac{N_{el,sat}}{2\pi h
b^2}\frac{e^{-\frac{(x-\tilde{r}_0)^2}{2\tilde{\sigma}^2}}}{F(1)}
\end{eqnarray}
and one of energy barrier which opposes the electron coming from
the wall, from (\ref{radialpotential}), is

\begin{eqnarray}\label{radialenergy}
\mathcal{E}(r)\,=\,-e\,V(r)\,=\,e\,V_0\biggl[g\ln
x+\frac{G(x)}{F(1)}\biggr]
\end{eqnarray}
and we report in Fig. (\ref{plot_density_energy}) their behavior.
It notes how the peak of the energy barrier corresponds to the
maximum concentration of electrons around the bunch charge. As
soon as the density of electrons tends to zero near the wall of
chamber also energy barrier tends to zero.


We conclude highlighting the main outcomes of this paper. The
quadratic map coefficient $\beta$ is analytically derived for the
evolution of an electron cloud density. The expression is in an
acceptable agreement when compared with results obtained after
ECLOUD simulations without magnetic field. The analysis is useful
to determine safe regions in parameter space where an accelerator
can be operated without creating electron clouds.

\end{document}